# Relativistic electron model in the outer radiation belt using a neural network approach

Xiangning Chu[1], Donglai Ma[2], Jacob Bortnik[2], W. Kent Tobiska[3], Alfredo Cruz[3], S. Dave Bouwer[3], Hong Zhao[4], Qianli Ma[2,5], Kun Zhang[6], Daniel N. Baker[1], Xinlin Li[1], Harlan Spence[7], Geoff Reeves[8]

[1] Laboratory for Atmospheric and Space Physics, University of Colorado Boulder, Boulder, CO, USA

[2] Department of Atmospheric and Oceanic Sciences, University of California, Los Angeles, California, USA

[3] Space Environment Technologies, Pacific Palisades, CA, USA

[4] Department of Physics, Auburn University, AL, USA

[5] Center for Space Physics, Boston University, Boston, MA, USA

[6] Space Science Institute, Boulder, CO, USA

[7] Institute for the Study of Earth, Oceans and Space, University of New Hampshire, Durham, NH, USA

[8] Space Science and Applications Group, Los Alamos National Lab, Los Alamos, NM, USA

Corresponding author: Xiangning Chu (chuxiangning@gmail.com)

**Key Points:**

- A neural network model was developed to forecast relativistic electron fluxes with energies > 1.8 MeV in the outer radiation belt (ORIENT-R)

- The ORIENT-R model reproduces the relativistic electron fluxes with high out-of-sample accuracy: Pearson $r$~0.95, uncertainty of a factor of ~2

- The ORIENT-R model reproduces short- and long-term dynamics of the outer radiation belt such as transport, acceleration, decay, and dropouts

**Abstract**

We present a machine-learning-based model of relativistic electron fluxes >1.8 MeV using a neural network approach in the Earth's outer radiation belt. The Outer RadIation belt Electron Neural net model for Relativistic electrons (ORIENT-R) uses only solar wind conditions and geomagnetic indices as input. For the first time, we show that the state of the outer radiation belt can be determined using only solar wind conditions and geomagnetic indices, without any initial and boundary conditions. The most important features for determining outer radiation belt dynamics are found to be AL, solar wind flow speed and density, and SYM-H indices. ORIENT-R reproduces out-of-sample relativistic electron fluxes with a correlation coefficient of 0.95 and an uncertainty factor of ~2. ORIENT-R reproduces radiation belt dynamics during an out-of-sample geomagnetic storm with good agreement to the observations. In addition, ORIENT-R was run for a completely out-of-sample period between March 2018 and October 2019 when the AL index ended and was replaced with the predicted AL index ([lasp.colorado.edu/home/personnel/xinlin.li](lasp.colorado.edu/home/personnel/xinlin.li)). It reproduces electron fluxes with a correlation coefficient of 0.92 and an out-of-sample uncertainty factor of ~3. Furthermore, ORIENT-R captured the trend in the electron fluxes from low-earth-orbit (LEO) SAMPEX, which is a completely out-of-sample dataset both temporally and spatially. In sum, the ORIENT-R model can reproduce transport, acceleration, decay, and dropouts of the outer radiation belt anywhere from short timescales (i.e., geomagnetic storms) and very long timescales (i.e., solar cycle) variations.

**Plain Language Summary**

The Earth's radiation belts consist of energetic particles. During periods of intense space weather, the energy and density of radiation belt particles can increase significantly and pose a danger to astronauts, spacecraft, and even technologies on the ground. This study presents a machine-learning-based model (ORIENT-R) that calculates the energetic radiation belt electron fluxes. It uses solar wind observations and geomagnetic activity indices as input, without the need for boundary conditions such as other satellite measurements. The ORIENT-R model can determine the energetic electrons with a high Pearson correlation of 0.95 and a small uncertainty factor of ~2. Furthermore, even when geomagnetic index AL is not available, the ORIENT-R model can still estimate with high accuracy of Pearson correlation of 0.92 and a small uncertainty factor of ~3. Thus, the ORIENT-R model has great value and wide application in the space physics community and space weather industry.

# 1 Introduction

The Earth's radiation belts consist of electrons that range in energy from hundreds of keV to multiple MeV (Baker et al., 2017; Thorne, 2010), and represent the first major discovery of the space age (Van Allen et al., 1958, 1959). Different particle species and energy ranges, most notably the electrons in the outer radiation belt, behave quite differently with respect to one another since different physical processes dominate their dynamics. Outer radiation belt electron fluxes are highly variable during geomagnetic storms due to a competition between various loss and acceleration processes (Bortnik et al., 2007; Li et al., 2006; Reeves et al., 2003; Turner et al., 2013; Xiao et al., 2009). Energetic plasma sheet electrons transported into the inner magnetosphere during substorms or periods of enhanced convection provide source electrons for exciting plasma waves, and seed electrons, which can be further energized to highly relativistic energies (Baker et al., 2008; Horne et al., 2003; Jaynes et al., 2015). During dynamic intervals, such as geomagnetic storms, both the source and loss processes controlling the radiation belts can operate on relatively short timescales (~1 day or less) and result in a net loss, acceleration, or no change to the pre-existing fluxes (Reeves et al., 2003). During non-storm periods, the balance of longer-timescale acceleration and loss processes (e.g., radial diffusion and pitch angle scattering) determines the radiation belts' overall configuration, which is often observed to gradually diffuse in the radial direction and decay exponentially in the period following rapid enhancements (Baker et al., 2014; Ma et al., 2015; Reeves et al., 2016). The radiation belts are a known hazard to satellites, and can cause deep dielectric charging, spurious signals, and total-dose-related degradation (e.g., Baker et al., 2016; Choi et al., 2011). Thus, developing a useful model to describe the trapped electron flux has long been a challenging yet important task.

A number of approaches have been used to specify the relativistic electron fluxes in the inner magnetosphere. The empirical models AE8 and AE9 (Ginet et al., 2013; Sawyer et al., 1976; Vampola, 1996), which cover the inner magnetosphere, is a statistically averaged model reconstructing long-term (solar cycle timescale) variations rather than specific events. Other models mostly focused on the relativistic electrons at geostationary orbit (GEO) or medium Earth orbit (MEO) orbit, which are only applicable at a fixed radial distance. Therefore, these models do not cover the outer radiation belt in terms of L shell and magnetic latitude. Techniques used include linear prediction filters (LPF) (Baker et al., 1990; Statistical Asynchronous Regression (SAR) method in O'Brien et al., 2001a, 2001b ; Optimum solar wind coupling function in McPherron et al., 2015), empirical modeling (Li et al., 2001; O'Brien, 2003; Xiao et al., 2009), neural networks (Fukata et al., 2002; O'Brien and McPherron, 2003; Koons et al., 1991; Ling et al., 2010; Shin et al., 2016; Zhang et al., 2020a), NARMAX (Balikhin et al., 2011; Boynton et al., 2013, 2015), and light gradient boosting algorithms (MERLIN in Smirnov et al., 2020). As the space physics community achieves a better understanding of the transport, acceleration, and loss mechanisms governing the evolution of the radiation belt, first-principle physics-based models (Salammbo model in Beutier et al., 1995a, b; BAS radiation belt model in Glauert et al., 2014; UCLA 3-D diffusion code in Ma et al., 2015; DREAM model in Reeves et al., 2012; VERB code in Subbotin et al., 2009; DREAM3D diffusion model in Tu et al., 2013) have been used to provide a reconstructions of MeV radiation belt electrons.

Bortnik et al. (2016a) proposed using a unified approach for reconstructing the global, time-varying distribution of any physical quantity sparsely sampled at various locations within the magnetosphere at different times using an artificial neural network (ANN). This approach has been applied to reconstruct the plasma density, for specification of the whistler-mode chorus and hiss waves, and for modeling the energetic electron fluxes (>1.8 MeV) in the inner magnetosphere (Bortnik et al., 2018; Chu et al., 2017a, 2017b). These models are driven by the time series of solar wind conditions and geomagnetic indices. The application of machine learning techniques in space physics has advanced significantly over the past few years due to many factors that include enormously increased volumes of data, significantly improved algorithms implemented in free and powerful libraries, and substantially more powerful computation hardware (Camporeale, 2019). More recently, the 1 MeV electron fluxes in the outer radiation belt have been modeled with boundary condition of electron fluxes from a combination of POES satellite data collected at low earth orbit and LANL-01A at GEO orbit with correlation coefficient with a prediction efficiency between 0.61 to 0.92 at different $L$ shells (PreMevE model in Chen et al., 2019; Lima et al., 2020); or only the electron fluxes from LEO POES satellite using ANNs (SHELLS model in Claudepierre et al., 2020c). Note that the coefficient of determination $r^2$ is high (0.83) between the POES >700 keV electron fluxes and the 1 MeV electron fluxes observed by Van Allen Probes (see Figure 1 in Claudepierre et al., 2020c), since the POES electron fluxes are precipitated from the equatorial electron fluxes measured by Van Allen Probes. Therefore, the previous models are essentially using precipitated electron fluxes to reconstruct or predict the equatorial electron fluxes. It is unresolved whether the radiation belt dynamics can be addressed or reconstructed using only solar wind parameters and geomagnetic indices, without the use of electron fluxes as initial and boundary conditions.

We present an ANN model of outer radiation belt relativistic electron fluxes covering a larger range of energy levels between 1.8 MeV to 6.5 MeV and $2.6 < L < 6.5$ in the inner magnetosphere based on Van Allen Probes data as described in Section 2. For the first time, we show that the variation of the outer radiation belt can be largely determined by solar wind driving and geomagnetic activity. The neural network model takes these parameters as inputs without relying on any initial and boundary conditions from other satellite data, making this model a unique specification tool. We demonstrate that the neural network model can capture the long-term (solar cycle timescale) and short-term (geomagnetic storm timescale) dynamics to reproduce important features of the outer radiation belt under various geomagnetic driving conditions.

## 2 Data Description

In this study, the relativistic electron fluxes in the Earth's outer radiation belt are modeled using the ORIENT-R model. The primary dataset consists of spin-averaged energetic electron flux measurements obtained from the Relativistic Electron Proton Telescope (REPT) instrument (Baker et al., 2012) onboard the Van Allen Probes (RBSP) (Mauk et al., 2012). The identically instrumented twin-spacecraft mission has a highly elliptical low-inclination orbit with an apogee of ~ 6 $R_E$ and a perigee of ~ 600 km. The relativistic electron fluxes are measured in the energy range ~1.5- 20 MeV and the REPT data from the RBSP-A and RBSP-B spacecraft are cross-calibrated to within 1% of one another. Thus, we used both the REPT-A and REPT-B data as a whole. The complete set of the REPT measurements spans from September 1, 2012, to the end of the mission (October 14, 2019 for RBSP-A and July 16, 2019 for RBSP-B). The temporal resolution was reduced to 1 min averages, and the L shell was restricted above 2.6 to focus on the outer radiation belt, which results in over 7.4 million data points in total.

The variation of the Earth's outer radiation belt is known to be driven by the solar wind (e.g., solar wind pressure via magnetopause shadowing, convection, and ultra-low frequency waves) and the resultant geomagnetic activity (e.g., whistler-mode chorus and hiss waves induced by the plasma injections from the magnetotail, as well as radial diffusion). Therefore, the solar wind conditions and geomagnetic indices obtained from the OMNI data set at 1 min resolution (https://omniweb.gsfc.nasa.gov/) are used as input parameters to the neural network model.

## 3 Methodology

### 3.1 Model description

In this study, the electron flux model is developed using a simple feedforward neural network consisting of three hidden layers. The architecture of the neural network for the electron flux model is similar to that used in previous studies, which succeeded in modeling global dynamic distributions of the plasma density, waves, and electron fluxes (Bortnik et al., 2016a, 2018; Chu et al., 2017a, 2017b). The input parameters (discussed in the next paragraph) used for each neuron are the products of the output of preceding nodes with their associated weights

$$z_j^l = f\left(\sum_{i=0}^{N-1} z_i^{l-1} w_{ij} + b_i\right),$$

where $i$ and $j$ denote the neuron number in the preceding and current layers, respectively, and $w_{ij}$ and $b_i$ are the weights and biases in the hidden layer. The output of each neuron in the hidden layers is calculated using a sigmoid activation function

$$f(z^l) = 1/(1 + \exp(-z^l)),$$

which is commonly used for ANN regression problems. The outputs of the preceding neurons are all used as the inputs of the neurons in the next layer, thus creating a fully-connected feedforward neural network. The electron flux model is trained using the Nesterov-accelerated Adaptive Moment Estimation (Nadam) optimizer to minimize the mean squared error (MSE) of the logarithm of the electron flux $\log_{10}$(flux). In order to avoid data leakage, we try to avoid splitting nearby data points into the training, validation, and test datasets. In that case, the model may memorize the data points in the training dataset, and use its memory to predict the nearby data points in the validation and test dataset. Therefore, we split the whole dataset is split into 30-day segments. Then, we randomly select 60% of these segments as the training set (~4.2 million data points), 20% as the validation set, and 20% as the test set. The training process continues to update the weights and biases using the training dataset, and its performance on the validation dataset is monitored. The training process stops when the MSE of the validation set stops improving for several consecutive epochs. We choose 15 epochs empirically, which is the same as Chu et al. (2017b). These early stopping criteria are widely used to avoid overfitting and ensure the generalizability of the electron flux model. In addition, the dropout technique is applied to each hidden layer to further improve the generalization of the ANN model by randomly dropping units from the neural network during training (Srivastava et al., 2014). The MSE of the test dataset, which the model has never seen, is used to indicate its complete out-of-sample performance (or predictive capability). Also, the electron flux data observed by both probes between February 22, 2017, and March 24, 2017, during which a geomagnetic storm took place, has been explicitly held out as an additional test dataset to quantify the out-of-sample performance of the electron flux model during geomagnetically active conditions.

The electron flux model's target is chosen to be the logarithm of the electron flux $\log_{10}(\text{flux})$ instead of the raw values of flux because the flux values can vary across several orders of magnitude, and the probability distribution of $\log_{10}(\text{flux})$ is much closer to a Gaussian distribution which is more amenable to the training procedure. The input parameters include the location of the measurements (i.e., $L$-shell), and the time series of geomagnetic indices Sym-H, AL, and solar wind velocity and proton density. The Sym-H index is affected by the ring current strength and indicates the phases of geomagnetic storms. The AL index, which is the lower envelope of the auroral electrojet indices, is an indicator of plasma injection from the magnetotail and is a proxy for substorms. These indices are obtained from the OMNI database. Note that the AL index is only available until February 28, 2018. Time series of these indices rather than instantaneous values are used as input because the outer radiation belt variation strongly depends on its preceding states. We used 2-hour averages of the Sym-H, AL, and solar wind speed and density for the preceding day and 12-hour averages for the preceding 30 days (including the end point). Thus, each data of electron fluxes is reconstructed using the ensemble of the time series of four OMNI parameters and three positional parameters of the Van Allen Probes (a total of 291 parameters). The rationale for the selection of these parameters is described in section 3.2.

### 3.2 Feature selection and hyperparameter optimization

The process of feature selection used here is primarily based on the strategy of adding the most informative predictors for the ORIENT-R model (Kuhn et al., 2013) and assessing its performance. First, the ORIENT-R model is trained using a time series of only one input parameter at a time from the OMNI database. We use modified stratified 9-fold cross-validation for the training process (https://scikit-learn.org/stable/modules/cross_validation.html). After looping through all the input parameters, we choose the AL index, which shows the best test performance among all input parameters. Second, the ORIENT-R model is trained using the time series of the AL index and one other input parameter from the OMNI database, where again, all (second) input parameters are tried and assessed in turn. Solar wind flow speed shows the best test performance among the rest of the input parameters. Third, we repeat the second step by adding the most informative input parameter for each iteration until the model performance on

(MSE) the test dataset increases by less than 0.001. Finally, we choose an optimal set of parameters: the AL index, SYM-H index, solar wind flow speed, and proton density.

To determine the optimal length of the time series for input parameters, we investigated the model performance at different time lengths. It should be noted that different input parameters may have different optimal lengths of lookback windows. However, a complete grid search for possible combinations of lookback windows is computation impossible. In addition, since a few techniques (e.g., dropout, regularization, early stopping) are used to avoid overfitting, the contribution from unimportant time delays of input parameters are minimized. This is further demonstrated by the fact that our model has very good performance on out-of-sample test datasets (further discussed in sections 4 and 5). Figure 1 shows the performance of the ORIENT-R model using different lengths of input parameters. The blue error bar shows the median and standard deviation of the test performance (MSE) based on modified stratified 9-fold cross-validation, while the red error bar shows those of the Pearson correlation coefficients. The best performance is found when a 30-day window of input parameters is used. This 30-day lookback window is consistent with the estimated decay timescale (~20 days) of the electron flux variations at 2-6 MeV energies in the outer radiation belt (Ripoll et al., 2014, 2020; Claudepierre et al., 2020a, 2020b; Thorne et al., 2013a), which is the longest timescale that needs to be resolved when the outer belt is simply decaying without any other (shorter scale) dynamics present. It can also be attribute to the 27-day synodic solar rotation, which was found between the relativistic electron fluxes and the upstream solar wind speed (Miyoshi et al., 2004; Wing et al., 2018, and references therein).

The hyperparameters of the ORIENT-R model, including the number of neurons in each hidden layer and the dropout rates, are optimized using a Tree-structured Parzen estimator algorithm (Bergstra et al., 2011, 2013) implemented in Optuna (Akiba et al., 2019). We use modified stratified 5-fold cross-validation, and the model yields the best performance on the test dataset based is chosen. The final ANN model has three hidden layers with 386, 183, and 15 neurons, respectively.

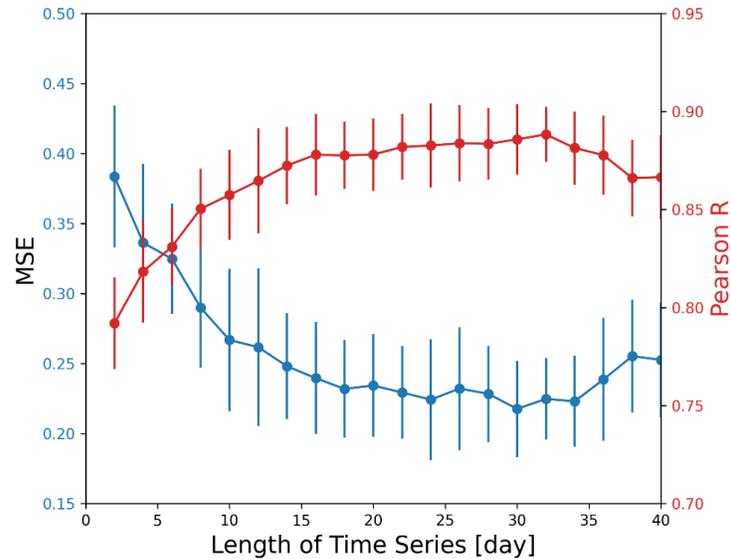

**Figure 1**. Model performance on the test dataset using different lengths of time series as input parameters. The blue line and error bars show the median and standard deviation of the MSE, and the red line and error bars show the Pearson correlation coefficient.

**4 Model application**

The ORIENT-R model was applied to the one-month period between February 22, 2017, and March 24, 2017, during which a moderate storm occurred. Note that the entire period containing the storm was all held out as part of the test dataset, which is out-of-sample and not used during the training process. Therefore, this case is representative of the predictive capability of the model on an out-of-sample dataset.

The background solar wind conditions and geomagnetic activity are shown in Figures 2a and 2b. A sudden increase in solar wind speed and density occurred on March 1, 2017. A geomagnetic storm started around 0900 UT when the IMF Bz turned southward. The Sym-H index decreased to a minimum value of -74 nT, then slowly recovered over the next 20 days until March 16, 2017, when another weak storm took place. The recovery of the Sym-H index was not monotonic, likely due to additional substorm injections as indicated by the AL index.

Figures 2c and 2d show the comparison between the observed (red) and modeled (blue) electron fluxes along the RBSP-A and RBSP-B trajectories. The ORIENT-R model is seen to reproduce the electron fluxes along the RBSP trajectories well, as the lines are very close and overlap with to each other. The correlation coefficient during the out-of-sample storm event is excellent, with a correlation coefficient $r$ = 0.96 and a root-mean-square deviation (RMSE) of ~0.34. The out-of-sample performance metrics are consistent with the test performance (r ~ 0.96 and RMSE~0.35), as shown in Figures 3 and 4. The result suggests that the ORIENT-R model can estimate out-of-sample electron fluxes within a factor of ~2 ($10^{0.35}$) for a sequentially held-out dataset or future dataset beyond the Van Allen Probes era.

Figures 2e and 2f show the comparison between the observed and reconstructed 1.8 MeV electron fluxes as a function of L shell and time. Figure 2g shows the modeled electron flux of 1.8 MeV on the equatorial plane (assuming magnetic latitude of 0°) at midnight (MLT=0). The ORIENT-R model reproduces the overall variation of relativistic electrons and many important features of the outer radiation belt, including all the physical processes involved resulting in magnetopause shadowing, local acceleration, radial diffusion, and decay due to pitch angle scattering. Firstly, the ORIENT-R model can reproduce the dropout of the outer radiation belt observed at 1.8 MeV that occurred on 1-2 March 2017, most probably due to magnetopause shadowing and subsequent outward radial diffusion when the solar wind pressure reached a maximum, and the relativistic electrons were lost to the dayside magnetopause. Interestingly, another weaker dropout event occurred on February 23, 2017, due to another solar wind pressure enhancement, which was also captured by the ORIENT-R model. Secondly, the ORIENT-R model can reproduce the local acceleration during the early recovery phase of the geomagnetic storm. The relativistic electron flux enhancement first occurs at the heart of the acceleration region, around L~5.0, which is consistent with the observation and previous theory of local acceleration (Thorne et al., 2013b). The reproduced enhancement first occurs in the 1.8 MeV electron fluxes and later in the higher energies due to their slower acceleration (not shown), consistent with observations. Third, the ORIENT-R model can reproduce the effects of radial diffusion during the storm. The modeled electron flux shows that, after the electron flux increase at L~5 during the early recovery phase, the enhancement proceeded toward both higher and lower L shells during 4-12 March period, most likely due to radial diffusion. Fourth, the ORIENT-R model can reconstruct the level and location of the peak flux at different energies throughout this geomagnetic storm event. The peak flux of 1.8 MeV electrons is well reproduced, as shown in Figures 2c and 2d and the comparison between Figures 2e and 2f. The reconstructed peak location in the 1.8 MeV electron fluxes gradually moved toward lower L shells (by ~$0.5L$) during the geomagnetic storm recovery phase, probably due to inward radial diffusion (Ma et al., 2015). Fifth, the decay of the relativistic electrons fluxes is well captured as the electron fluxes slowly decrease at both heart and edges of the outer radiation belt during the late recovery phase. The decay time of the electron flux is also shorter at larger L shells, consistent with previous observation and theory on electron lifetimes (Claudepierre et al., 2020a, 2020b). Finally, the electron fluxes exhibit additional sporadic enhancements during the recovery phase, which seems to be related to the enhanced geomagnetic activity (e.g., between 8-9 March 2017). In summary, the ORIENT-R model reconstructed the evolution of relativistic electron fluxes in the outer radiation belt with good performance and captured the key features and details of the flux dynamics during a geomagnetic storm.

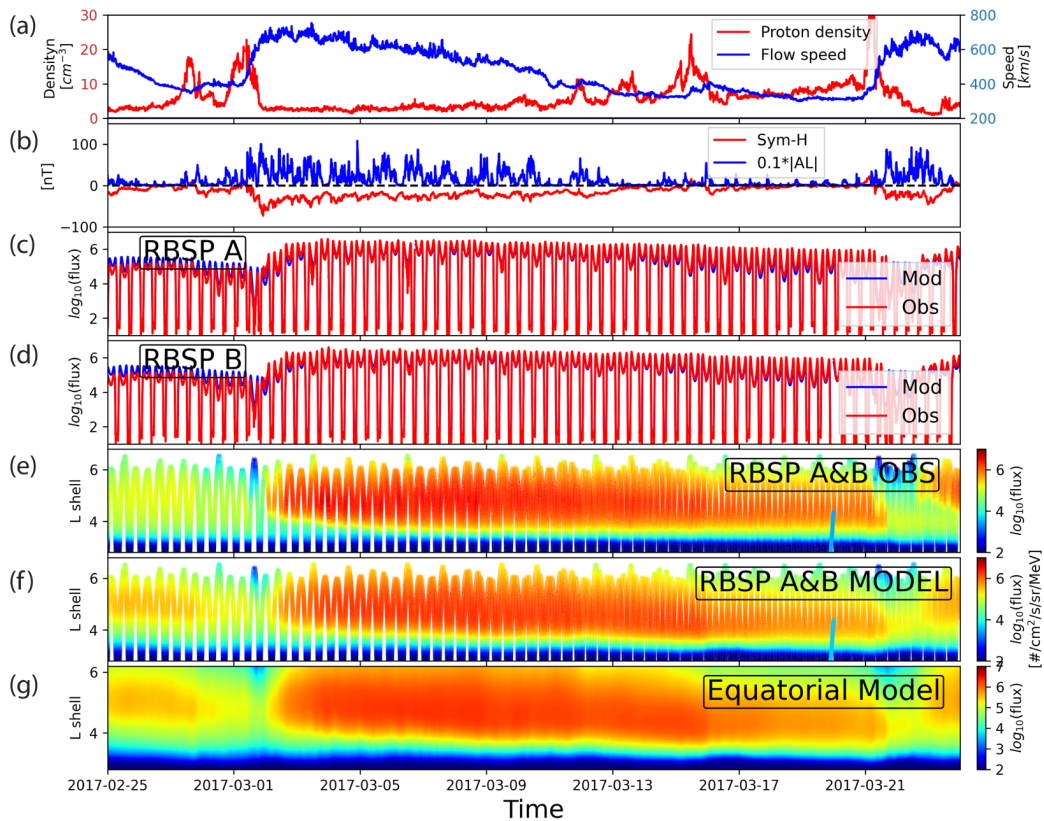

**Figure 2**. An application of the ORIENT-R model during the month-long period between February 20, 2017, and March 20, 2017, which was held out for test purposes. (a) The solar wind flow speed and density; (b) geomagnetic indices Sym-H and AL; (c-d) the observed and modeled 1.8 MeV electron fluxes along the trajectories of Van Allen Probes A and B; (e-f) the observed

and modeled 1.8 MeV electron fluxes as a function of L shell and time; (g) the modeled 1.8 MeV electron fluxes on the equatorial plane (MLAT=0°) at midnight (MLT=0).

## 5 Model performance

The correlation between the observed and modeled electron fluxes of 1.8 MeV is shown in the four panels in Figure 3 for the whole dataset, as well as the training, validation, and test datasets separately. The colors represent the probability density of occurrences in each bin. The red dashed line represents the diagonal line (y = x) where the model perfectly calculated the observed electron fluxes. The majority of the observation-model pairs are distributed around the diagonal line, indicating that the ORIENT-R model reproduces the observations accurately without over/under estimation. Note that the Pearson correlation coefficient and the square root of the coefficient of determination $r^2$ are very close (within 0.0001) since the observation-model pairs are distributed along the diagonal line. The Pearson correlation coefficients $r$ of all four datasets are ~0.95, which means that the neural network can explain $r^2$ = 90% of the observed variability in the electron fluxes. The RMSE on the test dataset of $\log_{10}$(flux) is 0.351, which translates to an uncertainty of a factor of 2.2 ($10^{0.351}$). Thus, the ORIENT-R model is expected to estimate out-of-sample observations within a factor of ~2. The model performance on the sequential test dataset in March 2017 is 0.340. Therefore, the ORIENT-R model is expected to have an excellent ability to make out-of-sample estimations.

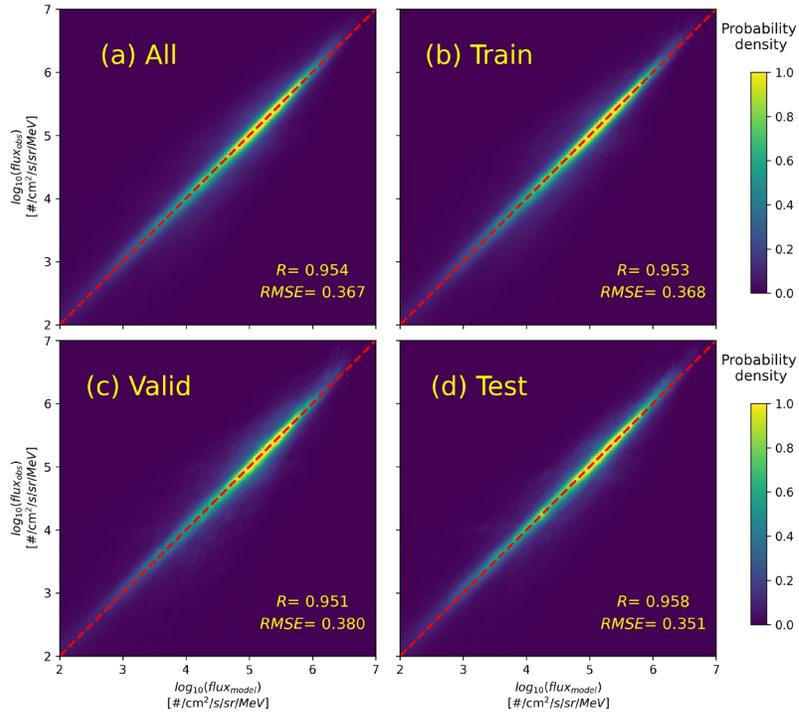

**Figure 3**. Two-dimensional probability density of the observed 1.8 MeV electron fluxes and those calculated by the ORIENT-R model for the four datasets (all, training, validation, and test). The red dashed line is the diagonal line that indicates perfect agreement (y=x) between the observations and model results. The Pearson correlation coefficients *r* and RMSE are shown at the right bottom corners.

Figure 4 shows the probability density of the errors as a function of the *L* shell for the four datasets: all, training, validation, and test. The error is defined as the difference between the logarithms of the observed electron fluxes and the modeled values $log_{10}(\text{flux}_{obs}) - log_{10}(\text{flux}_{model})$. The mean errors are close to zero, indicating that the model does not have significant bias. The RMSE indicated by the error bars is small at the heart of the outer radiation belt between L~4-6. For instance, an RMSE of 0.3 at L=4.5 on the test dataset translates to an uncertainty of a factor of ~2.0 ($10^{0.3}$). The RMSE is relatively larger in the slot region below

L~4.0, where flux values are low. For instance, an RMSE of 0.60 at L~3.0 translates to an uncertainty of a factor of 4. The probability density of the observed electron fluxes versus the $L$ shell is shown in Figure 4e, illustrating the variability of the outer radiation belt, which changes by a few orders of magnitude during geomagnetic storms. The black line is the mean values of observed fluxes, and the error bars are the RMSE of the ORIENT-R model on the whole dataset. Figure 4e demonstrates that ORIENT-R's uncertainty is much smaller than the level and variability of the electron fluxes. In sum, the ORIENT-R model has outstanding performance in the outer radiation belt.

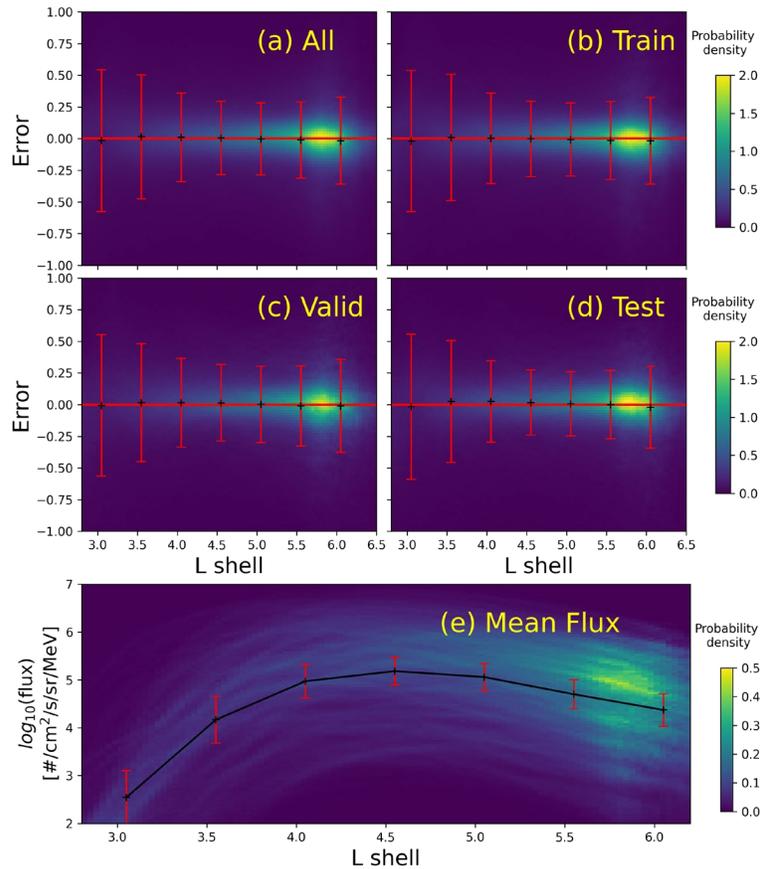

**Figure 4**. The probability density of the errors as a function of $L$ shell for the four datasets: (a) all, (b) training, (c) validation, and (d) test. The error is defined as the difference between the logarithms of the observed electron fluxes and the modeled values: $log_{10}(flux_{obs}) - log_{10}(flux_{model})$. The error bars illustrate the mean errors and RMSE as a function of the $L$ shell for four models. Panel (f) shows the probability density of the observed electron fluxes from Van Allen Probes. The black line illustrates the mean electron fluxes in the outer radiation belt, and the error bars illustrate the RMSE of the ORIENT-R model.

The ORIENT-R model was applied to reconstruct the electron fluxes during the Van Allen Probe era along their trajectories. Figure 5 shows an overview of the neural network model results versus observations from September 2012 to March 2018 (when the AL index ended).

Figure 5 shows that the long-term variations in the outer radiation belt are successfully reproduced. The neural network model captured the variation of the fluxes during all the storms in the five-year period, including the minimum values of the electron fluxes during dropout events, the maximum of the electron fluxes during storms, the timing and $L$ shell of the rapid local acceleration, and the accurate magnitude of electron fluxes during each storm. The residuals in Figures 5c and 5f are mostly clustered around zero at the heart of the outer radiation belt at $3.5 < L < 6.5$. The peak electron fluxes are underestimated or overestimated during only a few storms, while the error is nevertheless relatively small compared to the typical 2-3 orders of magnitude (a factor of 100-1000) variation of the electron fluxes during geomagnetic storms. The model performance has reached a satisfactory level of accuracy to be able to study geomagnetic storms compared to existing statistical models. Note that the error was higher at the inner edge of the outer radiation belt (L<4) during a few specific time intervals (e.g., around January 2013, December 2013, June 2017, as well as September 2018 in Figure 6). The inner edge of the outer radiation belt reconstructed by the ORIENT-R model was offset by a few tenths in L shell during these times. The electron flux in the vicinity of the edge changed significantly from ~10 (about the noise level) in the slot region to ~1,000 in the radiation belt (see Figures 2c and 2d for reference). Therefore, it resulted in large errors near the edge,

$$log_{10}(flux_{obs}) - log_{10}(flux_{model}) = log_{10}(flux_{obs}/flux_{model}),$$

which depends on the ratio of the modeled and observed electron fluxes. However, the overall variation of the outer radiation belt was well captured.

The second point to note is that the solar cycle variation of relativistic electron fluxes is reproduced by the neural network model. The electron fluxes are moderate during the ascending phase (late 2012 - early 2013), extremely low during an unusually low sunspot number period (late 2013 – 2014), and high during the descending phase (2015 to 2018), as shown in Figures 5a and 5d. These long-term variations are reproduced as shown in Figures 1b and 5e. Interestingly, the ≥4.2 MeV electron fluxes are extremely low for the entire year of 2014 (Baker et al., 2019), when the solar wind speed Vsw values were continuously low in 2014. The low solar wind speed during the extra ascending phases in 2014 corresponds to lower geomagnetic activity and significantly fewer substorms (Hsu et al., 2012; McPherron et al., 2013; Chu et al., 2015). Thus, the extremely low flux is likely due to fewer substorms which produce fewer injections, and thus fewer seed electron, probably weaker chorus waves, and finally, less acceleration of relativistic electrons.

Third, the electron flux enhancement at extremely low L shell during strong storms is captured by the ORIENT-R model (e.g., March 2015, June 2015), indicating the model's capability to reconstruct electron flux enhancement during extremely disturbed geomagnetic storm events, even though the training set for such events is relatively small. In summary, the ORIENT-R model captured both the long-term (i.e., solar cycle) and short-term (i.e., geomagnetic storms) variations during the Van Allen Probes era.

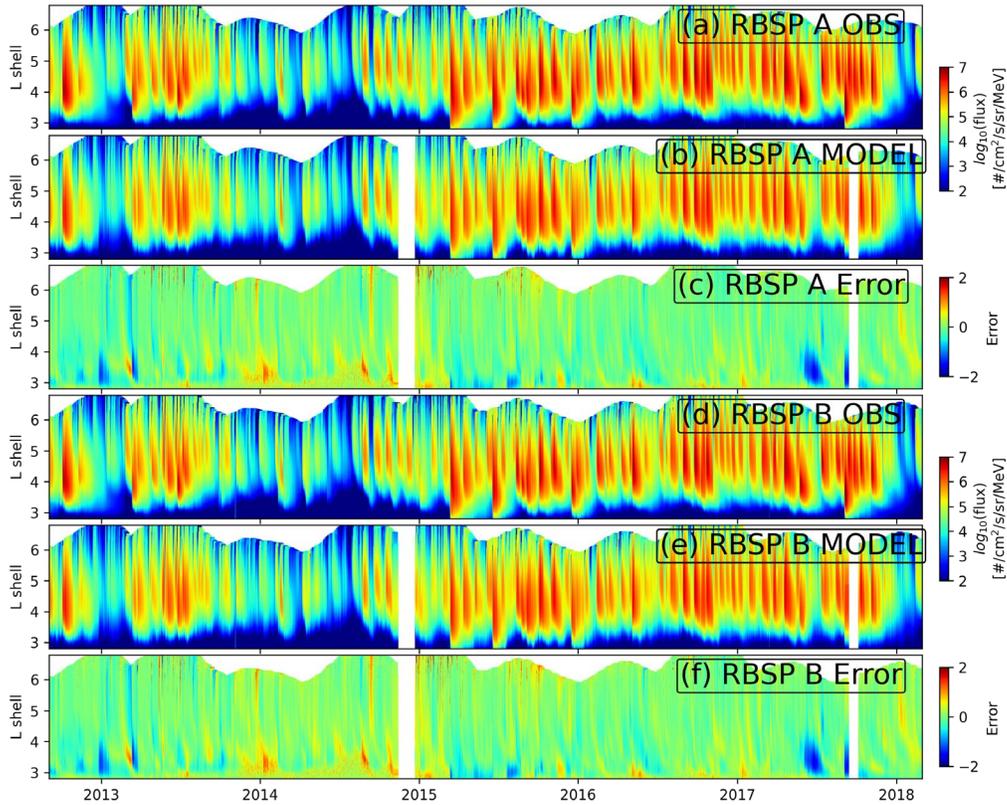

**Figure 5**. Overview of the ORIENT-R model reconstructions versus observations for Van Allen Probes from September 2012 to March 2018. The panels are: (a, d) observed 1.8 MeV electron fluxes from RBSP-A and RBSP-B; (b, e) the reconstructed electron fluxes from the ORIENT-R model along the trajectory of Van Allen Probes; (c, f) the differences between the observed and modeled electron fluxes, which are defined as $log_{10}(flux_{obs}) - log_{10}(flux_{model})$. The white vertical gaps in panels (b, c, e, f) are due to large data gaps in the OMNI database.

The ORIENT-R model is validated using the electron fluxes along Van Allen Probes' trajectories after March 2018 when the AL index ended. The electron fluxes during this period are not used in the training process. Therefore, the comparison shows the out-of-sample predictive ability of the ORIENT-R model. We used the predicted AL index using upstream solar wind measurements available at [lasp.colorado.edu/home/personnel/xinlin.li](lasp.colorado.edu/home/personnel/xinlin.li) (Li et al., 2007; Luo et al., 2013), which has a linear correlation coefficient of 0.846 and a prediction efficiency of 0.715 (meaning that 71.5% of the variations in the AL index are captured). Note that the AL index is the most important parameter for the ORIENT-R model (see section 3.2). Thus, the performance of the ORIENT-R model is expected to be degraded since the predicted AL index would introduce additional errors. The uncertainty in the electron fluxes caused by the uncertainties in the input parameters is outside the scope of the current study and will be quantitatively analyzed in a future study (e.g., the comprehensive analysis in Licata et al. (2020)).

Figure 6 shows the comparison between the observed and modeled 1.8 MeV electron fluxes along the trajectories of the Van Allen Probes between March 1, 2018 and October 13, 2019. The ORIENT-R model captures the overall dynamics of the outer radiation belt during all geomagnetic storms. Using the predicted AL index, the Pearson correlation coefficient $r$ during this period is ~0.92, which means that the ORIENT-R model can explain $r^2$ = ~85% of the observed variation in the electron fluxes. The RMSE on the electron fluxes $\log_{10}(\text{flux})$ is ~0.48, which translates to an uncertainty factor of 3.0 ($10^{0.48}$). We emphasize that the errors are contributed from both the error of the ORIENT-R model and the error in the predicted AL index. Nevertheless, the good correlation coefficient shows the out-of-sample predictive capability of the ORIENT-R model on the Van Allen Probes dataset.

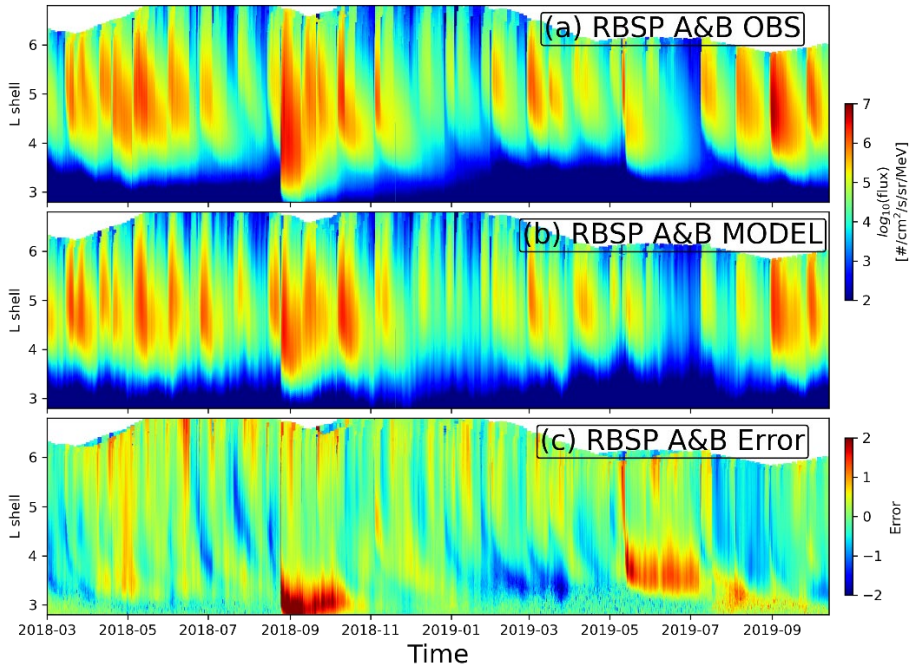

**Figure 6**. An out-of-sample application of the ORIENT-R model along Van Allen Probes' trajectories after March 2018 when the AL index ended. (a) The observed 1.8 MeV electron fluxes along the trajectories of Van Allen Probes. (b) The modeled 1.8 MeV flux from the ORIENT-R model. It used the predicted AL index (lasp.colorado.edu/home/personnel/xinlin.li) as input, while the other inputs are obtained from the OMNI database. The Pearson correlation coefficient is 0.92, and the RMSE is ~0.48. (c) the differences between the observed and modeled electron fluxes, which are defined as $log_{10}(flux_{obs}) - log_{10}(flux_{model})$.

The trapped electron fluxes reconstructed by the ORIENT-R model can be further validated in a completely out-of-sample fashion using the ~2 MeV electron fluxes measured by the Proton-Electron Telescope (PET) on the Solar Anomalous and Magnetospheric Particle Explorer (SAMPEX) satellite, which is located in a polar-orbiting low earth orbit (LEO, launched into a 690 × 510 km altitude and 82° inclination orbit in July of 1992, and decayed to ~490 × 410 km by 2009) (Baker et al., 1993; Cook et al., 1993; Li et al., 2020; Selesnick, 2015).

Since SAMPEX is located at LEO orbit, it measures electrons with a very small range of equatorial pitch angles very close to the loss cone, in contrast to the Van Allen Probes which observed the full equatorial pitch angle distribution, and thus the two datasets are closely correlated but not identical to each other, so the comparison is only qualitative (Kanekal et al., 2001; Li et al., 2013, 2017; Zhang et al., 2020b). Figure 7 compares the electron fluxes measured by the LEO SAMPEX satellite and the equatorial electron fluxes reconstructed by the ORIENT-R model in 2004. The figures of comparison from 1995 to 2009 are available at https://doi.org/10.5281/zenodo.5519666. Seven geomagnetic storms (SYM-H<-100 nT) took place in 2004. The relativistic electron fluxes measured by SAMPEX at LEO orbit show dynamic features of the outer radiation belt during these geomagnetic storms. The modeled equatorial electron fluxes show a similar trend to the SAMPEX observations. In addition, the inner edge of the outer radiation belt from SAMPEX observation seems well correlated with that of the ORIENT-R model, even during the large storms in July and November 2004. Note that the ORIENT-R model was trained using Van Allen Probes measurements obtained from a different region (equatorial v.s. highly inclined LEO) at different time ranges (after late 2012 vs. in 2004). Therefore, such a good correlation demonstrates the out-of-sample predictive capability of the ORIENT-R model on past and future datasets.

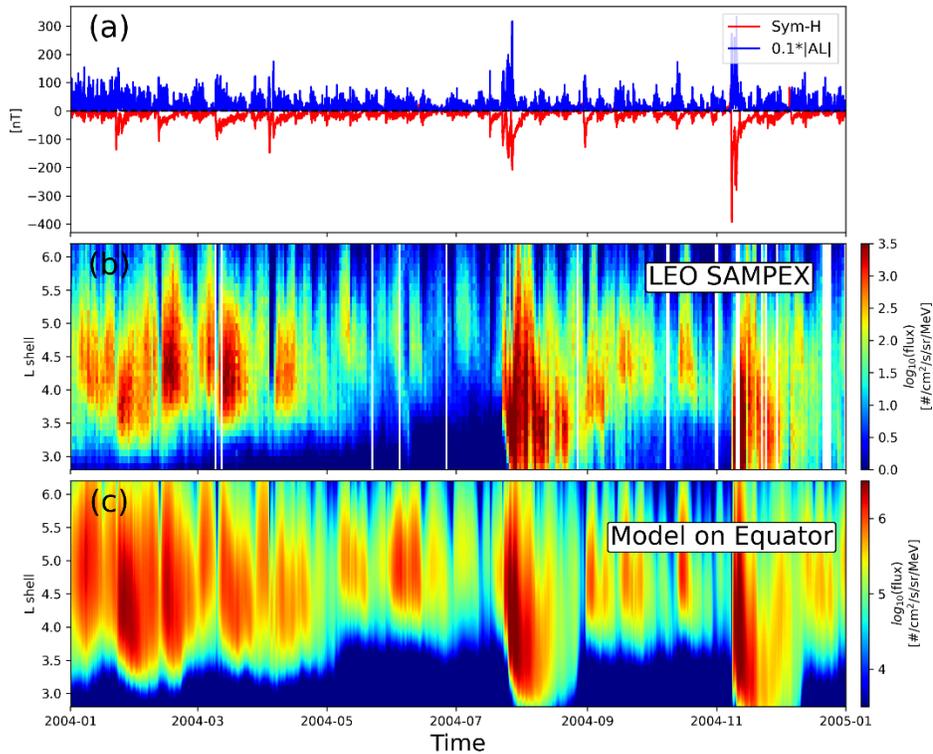

**Figure 7**. Comparison between SAMPEX observed electron fluxes at LEO orbit and the modeled equatorial electron fluxes obtained from the ORIENT-R model in 2004. (a) Geomagnetic indices Sym-H and AL; (b) half-day window-averaged and *L*-binned (*L* bin: 0.1) electron fluxes of ~ 2 MeV measured by SAMPEX; (c) reconstructed electron fluxes of ~1.8 MeV on the equatorial plane at midnight from the ORIENT-R model. The white gaps in the SAMPEX observations are due to a lack of observations.

Figure 8 shows the Pearson correlation coefficients on the test dataset versus energy channels. The Pearson correlation coefficient is highest (0.958) for the first energy channel (1.8 MeV), decreases as the energy increases, and reaches 0.807 for the 7$^{th}$ energy channel (6.2 MeV). The model performance is better at lower energies because the electron fluxes are much higher, and there are more events at lower energies. On the other hand, it is difficult for the electrons to be accelerated to higher energies; therefore, there are fewer events at higher energies and significantly more observations of low fluxes near the noise level (see Figure 1 in Baker et al., 2019). Thus, the model performance is degraded by the large number of low electron fluxes due to the imbalance of the dataset. Imbalanced regression techniques will be carried out in future studies to achieve better model performance at high energy channels.

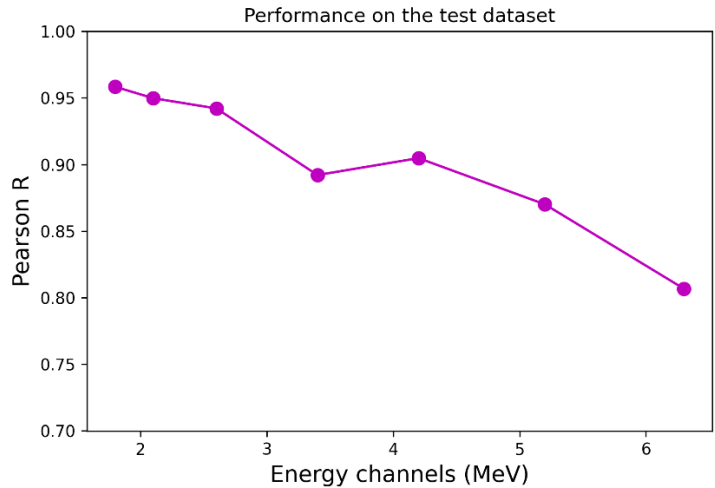

**Figure 8**. Pearson correlation coefficients on the test dataset versus electron energies.

**6 Discussion and Summary**

We present a neural network-based model of relativistic electron fluxes in the outer radiation belt, which utilizes a combination of the time history of geomagnetic indices and solar wind parameters as inputs to reconstruct the electron fluxes of relativistic energies in the inner magnetosphere. It should be noted that the ORIENT-R model does not rely on any boundary conditions such as particle observations from other satellites at all, but is purely driven by geomagnetic indices and solar wind parameters. The model has been trained and validated using the six years of electron flux data from the REPT instrument onboard Van Allen Probes. The model performance was demonstrated on the test dataset and an out-of-sample dataset of a geomagnetic storm in March 2017. The Pearson correlation coefficient is high (accounting for

~90% of the variations), and the modeled electron fluxes have uncertainty within a factor of ~2. The model can reproduce the transport, acceleration, decay, and dropout of the outer radiation belt electron fluxes during short timescales (i.e., geomagnetic storms) and long timescales (i.e., solar cycle) variations.

The ORIENT-R model has great value and wide application in the space physics community and space weather industry. First, this is the first machine learning-based model for reconstructing > 2 MeV electron flux of the outer radiation belt. Second, the model uses only solar wind parameters and geomagnetic indices and does not rely on any boundary conditions, e.g., other in situ particle data resources. It suggests that the radiation belt dynamics can be determined reasonably well using the time history of solar wind drivers and geomagnetic indices, without the need of initial condition and boundary conditions. This ML-based model is different from numerical simulations which usually rely on the accuracy of the initial and boundary conditions. The ORIENT-R model can reconstruct historical events, nowcast, and forecast future events as long as the model input parameters are available. To reconstruct historical events, the OMNI dataset is an ideal source to obtain the input parameters. For nowcasting, the input parameters can be obtained from the following sources: solar wind parameters from DSCOVR mission, DST index, and AL index predicted from upstream solar wind (Li et al., 2007; Luo et al., 2013), a neural network-based algorithm (Bala et al., 2012), and an optimum solar wind coupling function (McPherron et al., 2015). For forecasting up to six days into the future, the input parameters can be obtained from model predictions such as Anemomilos (Tobiska et al., 2013), Rice DST (Bala & Reiff, 2012), and ENLIL-WSA (Odstrcil, 2003). The ORIENT-R model has also been operationally deployed in real-time for usage by space-weather customers through a partnership with Space Environment Technologies

(sol.spaceenvironment.net/orbis_ops) with a nowcast. Note that the model was trained on the Van Allen Probes data, which started in late 2012 and ended operations in late 2019. Therefore, the 7-year dataset was taken during a relatively weak solar cycle, and not many strong or extreme storms occurred during the period in terms of radiation belt electron fluxes or geomagnetic activity (Meredith et al., 2015; Bernoux & Maget, 2020). However, future CubeSat missions launching soon will be deployed into a similar GTO orbit, which will further monitor the relativistic electron fluxes in the outer radiation belt (Blum et al., 2020). The ORIENT-R model can provide a critical prediction of the killer electrons for the satellites in orbit. In addition, the current model can be treated as a baseline, and periodically updated and improved with additional datasets from future missions, especially for strong and extreme storms during more active solar cycles.

The ML-based models can benefit physical insight in scientific research, especially when combined with physics-based numerical simulation. The ML models have the advantage of finding the non-linear relationship between the input parameters and the target parameter. Thus, the ML models usually have excellent capability to reconstruct or predict the target parameter. However, due to the black-box nature of artificial neural networks, it is difficult to extract the physics directly from the neural network models. Therefore, many efforts have been spent on interpretable machine learning in the last decade (Molnar, 2020). On the other hand, physics-based numerical simulations have included the essential physical mechanisms, especially in the quasilinear regime. Therefore, physics-based simulations are widely used in physics and have many advantages (e.g., easy to implement controlled parameter experiment by investigating one physical parameter's response to another physical parameter). However, the physics-based models usually require accurate knowledge of the physical processes involved and the initial and boundary conditions. For instance, the Fokker-Planck numerical simulation for radiation belt dynamics relies on accurate initial and boundary conditions of the electron fluxes, as well as the plasma wave distributions (and their temporal evolution) and total electron densities on a global scale to provide robust and accurate reconstructions (Ma et al., 2015). Therefore, a combination of ML models and physics-based numerical simulations has been carried out in physical discovery. For example, the ML-based models of the plasma waves and electron densities have been used to provide accurate initial and boundary conditions on a global scale to physics-based numerical simulations and reconstructed the radiation belt dynamics (e.g., Bortnik et al., 2016b; Ma et al., 2018). Thus, we believe that the ML-based models can be beneficial both as a model in itself, and combined with physics-based models.

In the future study, the phase space density distribution will be obtained using the electron fluxes from the ORIENT-R model to investigate radiation belt dynamics. In this study, we have used the spin-averaged electron fluxes categorized by the $L$ shell in our first study. Due to the adiabatic motion, the relativistic electron dynamics are more appropriately described using the phase space density distribution as a function of the three adiabatic invariants. Ideally, a well-trained neural network model can automatically incorporate the electron phase space density variation along the dimensions of the three adiabatic invariants. For example, the variation of electron drift path ($L^*$) during geomagnetic storms may be implicitly incorporated in our current model. From the Van Allen Probes observation, the electron phase space density data at different adiabatic invariants can be obtained from the pitch angle distribution of electron fluxes over a wide range of energies from both the REPT and MagEIS instruments. In addition, the electron fluxes from the two instruments need to be cross calibrated. A neural network-based model of the electron fluxes from the MagEIS data is developed (Ma et al., 2021). Using the phase space density distribution as the database from the neural network is closer to the underlying physics and may provide a more reliable model, which is left for future investigation.

**Acknowledgments**

The authors would like to thank the NASA SWO2R award 80NSSC19K0239 for their generous support for this project. XC would like to thank grant NASA ECIP 80NSSC19K0911 and NASA LWS 80NSSC20K0196. JB acknowledges support from the Defense Advanced Research Projects Agency under Department of the Interior award D19AC00009. We gratefully acknowledge the REPT team (www.rbsp-ect.lanl.gov), Van Allen Probe mission (rbspgway.jhuapl.edu), OMNI dataset (omniweb.gsfc.nasa.gov), DSCOVR (www.swpc.noaa.gov/products/real-time-solar-wind), predicted AL at RICE (mms.rice.edu/realtime), predicted AL from Xinlin Li ([lasp.colorado.edu/home/personnel/xinlin.li](lasp.colorado.edu/home/personnel/xinlin.li)), and SAMPEX data (http://www.srl.caltech.edu/sampex/DataCenter/data.html). The model files and the processed data are made available at [https://doi.org/10.5281/zenodo.5519666](https://doi.org/10.5281/zenodo.5519666). This work used Extreme Science and Engineering Discovery Environment (XSEDE) Bridges GPU at the PSC through allocation TG-PHY190033.